\documentclass[conference]{IEEEtran}

\usepackage{graphicx}
\usepackage{cite}
\usepackage[cmex10]{amsmath}
\usepackage{amsfonts}
\usepackage{amssymb}
\usepackage{subfigure}
\usepackage{color}
\usepackage{amsthm}
\usepackage{enumerate}
\usepackage{relsize}
\usepackage{url}

\newcommand{\bm}{\mathbf} 
\newcommand{\be}{\begin{equation}}
\newcommand{\ee}{\end{equation}}
\newcommand{\bse}{\begin{subequations}}
\newcommand{\ese}{\end{subequations}}
\newcommand{\bea}{\begin{eqnarray}}
\newcommand{\eea}{\end{eqnarray}}
\newcommand{\x}{{\bm x}}
\newcommand{\z}{{\bm z}}
\newcommand{\Z}{{\bm Z}}
\newcommand{\y}{{\bm y}}

\newcommand{\bD}{{\bf D}}

\newcommand{\h}{{\bf h}}
\newcommand{\bh}{{\bf h}}

\newcommand{\bv}{{\bf v}}

\newcommand{\by}{{\bf y}}

\newcommand{\eye}{{\bm I}}

\newcommand{\calZ}{{\cal Z}}

\newcommand{\Fourier}{\mbox{\boldmath$\cal F$}}
\theoremstyle{definition}

\graphicspath{{figures/}} 

\IEEEoverridecommandlockouts

\begin{document}

\title{Underlay Control Signaling for Ultra-Reliable Low-Latency IoT Communications \vspace{-0.0in}}

\author{ {\normalsize Amir Aminjavaheri$^\dag$, Ahmad RezazadehReyhani$^\dag$,  Ramon Khalona$^+$, Hussein Moradi$^\ddag$, and Behrouz Farhang-Boroujeny$^\dag$}\\
{\normalsize $^\dag$ECE Department, University of Utah, $^+$Independent Consultant, $^\ddag$Idaho National Laboratory}
\thanks{This manuscript has been authored by Battelle Energy Alliance, LLC under Contract No. DE-AC07-05ID14517 with the U.S. Department of Energy. The United States Government retains and the publisher, by accepting the paper for publication, acknowledges that the United States Government retains a nonexclusive, paid-up, irrevocable, world-wide license to publish or reproduce the published form of this manuscript, or allow others to do so, for United States Government purposes. STI Number: INL/CON-17-43672.} \vspace{-0.0in}}

\maketitle

\begin{abstract}
Future mobile networks not only envision enhancing the traditional link quality and data rates of mobile broad band (MBB) links, but also development of new control channels to meet the requirements of delay sensitive use cases. In particular, the need for ultra-reliable low-latency communications (URLLC) for many internet of things (IoT) users is greatly emphasized. In this paper, we present a novel spread spectrum waveform design that we propose for transmission of control signals to establish URLLC communications.  These control signals are transmitted over the spectral resources that belong to the MBB communications in the network, but at a level that minimally affects these data channels. The proposed waveform, although a direct sequence spread spectrum (DSSS) technique, is designed to take advantage of symbol synchronization available to the OFDM broad band communications in the network. This allows simple synchronization with the rest of the network. The proposed DSSS method can transmit single and multiple bits within each OFDM time frame and can serve many user equipment (UE) nodes simultaneously. 
\end{abstract}

\section{Introduction}
Future wireless networks will be heavily driven by internet of things (IoT) connectivity. It has been noted that many IoT use cases require ultra-reliable low-latency communications (URLLC) and with that regard variety of solutions have been proposed and are under study, \cite{Teyeb2017,Schulz2017}. While some solutions take the approach of grant-free communications \cite{3GPP-R1-1611223}, others emphasize the need for grant-based approaches \cite{3GPP-R1-1700828}.


In grant-free uplink (UL) transmissions, UEs transmit within a set of specified resource blocks of OFDM frames/slots without any explicit scheduling grants from the base station (BS). This clearly lowers the control signaling overhead and under certain conditions may minimize latency, \cite{3GPP-R1-1611223}. However, UL transmissions require the UEs to transmit within a given set of resources that are predefined and preassigned by the BS, with a certain periodicity. To avoid resource under-utilization, multiple UEs should share the same resources and operate in a contention-based manner.  Hence, collisions that affect both reliability and latency will be unavoidable. Clearly, reliability decreases and latency increases as the UE traffic increases, due to contention and, as a result, retransmissions. In other words, grant-free transmission may not be scalable, as the UE density increases in a network, \cite{3GPP-R1-1704309}. 

Considering the above observations, the industry is currently looking into grant-based solutions and are exploring methods that reduce latency and improve reliability. In particular, to reduce the round-trip delay between sending a scheduling request (SR) to acquire resources to transmit and receive a grant, LTE numerologies need to be modified. The 5G New Radio (NR) contributions, in 3GPP, have already introduced mini-slots of 2 OFDM symbols and have considered reducing the time duration of each OFDM symbol, e.g., see \cite{3GPP-R1-1701548} and \cite{3GPP-R1-1702489}. Despite these efforts, in grant-based methods also resources have to be allocated to UEs for sending their SRs. With limited resources available for this task, resources still have to be allocated with a certain periodicity, whose period increases with the number UEs within each cell. This, clearly, increases latency in UL transmissions.

Preliminary results in our 3GPP contribution \cite{3GPP-R1-1701612} have shown that a significant reduction in latency for URLLC services (on the order of 40\% or more with respect to LTE) can be achieved through the use of a scheduling method using SR signals that can be transmitted essentially asynchronously with the rest of the network, but over the same spectrum. In particular, in \cite{3GPP-R1-1701612}, we introduced a direct sequence spread spectrum (DSSS) signaling method for transmission of such SRs, and explored some of its fundamental features without getting involved in the mathematical details. Since, in our method, the SR communications are established over the full bandwidth of the network, but at a level that the generated SR signals minimally affect the normal communications in the network, we say SRs are transmitted over an underlay channel and thus are referred to as underlay SRs (USRs).

In this paper, we present the mathematical detail of the DSSS signaling method that has been introduced in \cite{3GPP-R1-1701612}. We also derive the mathematical equations that quantify the probability of correct detection of the transmitted USR signals as well as the probability of false alarms (i.e., a USR is detected when no USR has been transmitted). In addition, the mathematical equations that quantify the interference level introduced by the USRs to the normal communications in the network are derived.
 
 There are a number of important reasons that motivate the use of an underlay channel for sending scheduling requests in IoT applications.  The presence of a large number of devices/UEs in each cell translates to a large number of SR messages that these devices have to send to the BS. However, at any given time a very small number of the UEs (in the order of one or two) may send SRs simultaneously. In addition, each SR should contain a UE identifier signature plus a small number of control bits. For instance, some current discussion in 5G NR talk about transmission of a single bit per each SR. The DSSS signaling method that has been introduced in \cite{3GPP-R1-1701612} and is further studied in this paper makes use of spread spectrum techniques to embed the UE identifier signature in each USR. Moreover, since the number bits to be transmitted per SR is small, the spread spectrum technique is taken advantage of to keep the signal power spectral density at a minimal level, hence, minimally affect the normal communications in the network.


This paper is organized as follows. The details of CP-DSSS message construction and information recovery is presented in Section II. The receiver implementation is discussed in Section III. In Section IV, we show that the interference introduced by USR in a typical network can indeed be very low. Performance of the proposed USR signaling is analyzed in Section V.  Some numerical results are presented in Section VI, and the concluding remarks are made in Section VII.

\section{CP-DSSS Message Construction and Information Recovery}

As discussed above, a DSSS method is adopted for transmission of URS messages. Each USR message has a length of one OFDM symbol and is time synchronized with OFDM symbols that are used for normal communications in the network. Each USR message is designed to carry one or more information bits within a format whose details are presented below. Furthermore, a cyclic prefix (CP) is added to each USR message in the same form as in OFDM. However, as explained below, the presence of CP in USR messages, apart from making them synchronized with the rest of the communications in the network,  finds many more benefits that make the proposed signaling method unique for the application which is intended for.  

We note that DSSS with CP is not new. Bruhl et. al. \cite{Bruhl2002} suggested the addition of CP to DSSS signals to facilitate channel equalization in the frequency domain. The referred to their signaling method as CP-CDMA. Subsequent studies \cite{Vook2002a,Vook2002b} showed that the addition of CP to DSSS signals also facilitates beamforming and space multiplexing. In this paper, as discussed in the sequel, we add CP with a very different primary goal. 

To construct CP-DSSS messages, we use Zadoff-Chu (ZC) sequences, \cite{velazquez2016sequence}, as spreading gain vectors.  ZC sequences are popular in LTE and are commonly used for channel estimation. Our use of ZC sequences, here, follows different goals. We particularly make extensive use of the properties of the ZC sequences that follow. 

\subsection{ZC Sequences}

We begin with a ZC sequence of length $N$ represented by the column vector $\z_0$. A fundamental property of ZC sequences that becomes very useful in many applications, and more particularly in the application of interest here, is that $\z_0$ and its downwards circularly shifted versions $\z_1$, $\z_2$, $\cdots$, $\z_{N-1}$ form a set orthogonal basis vectors. If we assume that these vectors are normalized to a length of unity, this property implies that
\be\label{eq:ziHzj}
\z_i^{\rm H}\z_j=\delta_{ij} ,
\ee
where $\delta_{ij}$ is the Kronecker delta function.

Next, for $i\mbox{ and } j$ in the range of $0$ to $N-L$, and $L<N$, we define the pair of $N\times L$ circulant matrices 
\begin{align}
\Z_i&=[\z_i~\z_{i+1}~\cdots~\z_{i+L-1}] \\
\Z_j&=[\z_j~\z_{j+1}~\cdots~\z_{j+L-1}] ,
\end{align}
and note that (\ref{eq:ziHzj}) implies that when $|i-j|>L$,
\be
\Z_i^{\rm H}\Z_j={\bf 0}_L.
\ee
Also,
\be
\Z_i^{\rm H}\Z_i=\Z_j^{\rm H}\Z_j=\eye_L.
\ee
When $|i-j|\le L$, $\Z_i^{\rm H}\Z_j$ is a diagonal matrix with zeros at its first $|i-j|$ diagonal elements, and ones at its remaining diagonal elements. 

In the sequel, we make frequent reference to the set of ZC sequences $\z_0$, $\z_1$, $\cdots$, $\z_{N-1}$ that were defined above. We use $\calZ$ to denote this set, i.e., $\calZ=\{\z_0, \z_1, \cdots, \z_{N-1}\}$.

\subsection{CP-DSSS Message Construction}
We consider a network with $U$ active UEs. Each CP-DSSS (equivalently, USR) message, sent from one of the UEs, carries one reference bit and $K$ information bits. For a user $u$, we denote these bits by $b_{u,0}$, for the reference bit, and $b_{u,1}$ through $b_{u,K}$, for the information bits. The corresponding CP-DSSS message frame is thus expressed as
\be
\x_u=\sum_{k=0}^K b_{u,k}\z_{u,k} ,
\ee
where $\z_{u,k}$ are a set of spreading gain vectors that are chosen from the set $\calZ$ and are specific to user $u$. A CP is added to $\x_u$ before transmission.

The above process is repeated at all UE nodes that have a SR to transmit. At the BS, after removing CP, one will get
\be\label{eq:y}
\y=\sum_{u=1}^U\left(\sum_{k=0}^K b_{u,k}\Z_{u,k}\right)\bh_u +\bv ,
\ee
where $\bh_u$ is the size $L\times 1$ vector of channel impulse response between the $u$th UE and the BS, $\Z_{u,k}$ are a set of circulant matrices that are constructed in the same way as $\Z_i$ and $\Z_j$, and $\bv$ is the vector of noise plus interference from the normal communications in the network.  For those UEs that have not transmitted any USR, $b_{u,k}=0$, for all $k$. Moreover, to allow straightforward extraction of the reference and information bits $b_{u,k}$ from (\ref{eq:y}), the matrices $\Z_{u,k}$ are chosen such that the index numbers associated with their first columns are at least $L+1$ points apart from one another. Imposing this condition, implies that
\be\label{eq:ZuiHZuj}
\Z_{u,i}^{\rm H}\Z_{u,j}=\left\{\begin{array}{ll}
\eye_L, & i=j\\
{\bf 0}_L, & i\ne j.\end{array}\right.
\ee

One may also note that to avoid interference among the information bits from different USRs, the spreading gain vectors of all users should be selected such that 
\be\label{eq:Zu1iHZu2j}
\Z_{u_1,i}^{\rm H}\Z_{u_2,j}={\bf 0}_L ,
\ee
for any UE pair $u_1\ne u_2$ and all values of $i$ and $j$. It should be noted that in practice, the constraint (\ref{eq:Zu1iHZu2j}) is rather restrictive and thus may be violated at the cost of some interference among USRs that may be transmitted simultaneously.

\subsection{Information Recovery} \label{sec:recovery}
Pre-multiplying (\ref{eq:y}) with $\Z_{u,i}^{\rm H}$ and using (\ref{eq:ZuiHZuj}), we get
\be\label{eq:yui}
\y_{u,i}=b_{u,i}\bh_u+\bv_{u,i},\quad\mbox{for }i=0,1,2,\cdots,K
\ee 
where $\y_{u,i}=\Z_{u,i}^{\rm H}\y$ and $\bv_{u,i}=\Z_{u,i}^{\rm H}\bv$. Note that here we have ignored possible interference from another USR in case (\ref{eq:Zu1iHZu2j}) does not hold and an interfering USR has been sent simultaneously. We have ignored this interference because, as will be discussed later, such interference is unlikely to happen and with a good design it can be kept at a minimal level, when it happens.

With the reference bit $b_{u,0}$ set equal to $+1$, estimates of the information bits are obtained as
\be\label{eq:hatbui}
\hat b_{u,i}=\mbox{sgn}\left[\Re\left[\y_{u,0}^{\rm H}\y_{u,i}\right]\right],\quad\mbox{for }i=1,2,\cdots,K
\ee
where $\Re[\cdot]$ and $\mbox{sgn}[\cdot]$ denote the real part and the signum of the argument.
This result follows since $\y_{u,0}^{\rm H}\y_{u,i}=\left(\bh_u^{\rm H}\bh_u\right)b_{u,i}+w_{u,i}$,   $\left(\bh_u^{\rm H}\bh_u\right)$ is a real valued scaling factor, and $w_{u,i}=\left(\bh_u^{\rm H}\bv_{u,i}\right)b_{u,0}+\left(\bh_u^{\rm H}\bv_{u,0}\right)b_{u,i}+\left(\bv_{u,0}^{\rm H}\bv_{u,i}\right)$ is a zero-mean random variable that arises from channel noise and any interference from other communications, including other USRs that may be transmitted simultaneously.

Here, we note that due to the special construction of the CP-DSSS, the information recovery is done without estimating the channel impulse response. This greatly simplifies the receiver implementation and allows transmission of a SR within each OFDM symbol frame.

\subsection{CP-DSSS Signal Detection} \label{sec:detection}

The above information recovery steps assume the BS is aware that the $u$th UE has sent a USR. Hence, prior to information recovery, the BS has to take an action to identify which UEs have transmitted USRs, if any. 

To detect whether a USR message has been transmitted by the $u$th user, we recall the set of equations in (\ref{eq:yui}) and note that if no message has been transmitted $b_{u,i}=0$, for $i=0,1,2,\cdots,N-1$, otherwise, $b_{u,i}$ are a set of random binary numbers taking values of $\pm 1$. We, thus, form the set of random variables
\be\label{eq:cij(u)}
c_{i,j}^{(u)}=\left|\Re\left[\y_{u,i}^{\rm H}\y_{u,j}\right]\right| ,
\ee
for all pairs of $i\ne j$ in the range of $0$ to $K$.

Under the hypothesis $H_0$, when no USR message has been transmitted by the $u$th user, $c_{i,j}^{(u)}$ are a set of random variables of the form $\left|\Re\left[\bv_{u,i}^{\rm H}\bv_{u,j}\right]\right|$. These random variables are likely to be concentrated near zero. On the other hand, under  the hypothesis $H_1$, when a USR message has been transmitted by the $u$th user, $c_{i,j}^{(u)}$ are a set of random variables whose distribution is significantly different from the case of $H_0$. This is because of presence of the term $\left(\bh_u^{\rm H}\bh_u\right)$ which biases the results to some values that are significantly distanced from zero. Accordingly, the presence or absence of a particular USR can be examined by evaluating the $c_{i,j}^{(u)}$ values under the hypotheses $H_0$ and $H_1$. More details of the simultaneous detection of USR signals for all the UEs are discussed in the next section. The theoretical analysis that evaluates the distribution of the random variable $c_{i,j}^{(u)}$ under hypothesis $H_0$ and $H_1$ along with other details/properties of the proposed CP-DSSS are presented in Section~\ref{sec:Analysis}.


\section{Receiver Implementation}\label{sec:ReceiverImplementation}
At the receiver, the set of vectors $\y_{u,i}$ are calculated collectively by taking the following steps.
\begin{enumerate}
\item
Calculate the length $N$ vector
\be\label{eq:y'}
\y'=\Z \y ,
\ee
where $\Z=\left[\begin{array}{cccc} \z_0& \z_1 & \cdots & \z_{N-1}\end{array}\right]$ is and $N\times N$ circulant matrix.
\item
For $u=1,2,\cdots,U$ and $i=0,1,2,\cdots,K$, set $\y_{u,i}$ equal to $L$ consecutive elements of $\y'$ that begin at the index that matches the spreading gain vector $\z_{u,i}$.
\end{enumerate}

Once $\y_{u,i}$  are obtained, the set of decision variables $c_{i,j}^{(u)}$ are calculated, using (\ref{eq:cij(u)}), for each user and compared against the hypothesis distributions $H_0$ and $H_1$. Subsequently, for those users whose decision variables $c_{i,j}^{(u)}$ found to belong to the hypothesis $H_1$ distribution, the corresponding information bits may be extracted according to (\ref{eq:hatbui}). If the information bits are coded, the signum function may be removed from (\ref{eq:hatbui}) to get the corresponding soft information which will be then passed to a soft decoder.
 
In order to further reduce the complexity of the receiver, we note that the circulant property of the matrix $\Z$ can be benefited from for a low complexity implementation of (\ref{eq:y'}). Since $\Z$ is circulant, it can be expanded as
\be\label{eq:Z}
\Z=\Fourier^{-1}\bD\Fourier ,
\ee 
where $\Fourier$ is the DFT matrix and $\bD$ is a diagonal matrix. Using (\ref{eq:Z}), $\y'$ can be calculated by (i) taking the DFT of $\y$; (ii) multiplying the elements of the result by the diagonal elements of $\bD$;  (iii) taking the IDFT of the result. Assuming $N$ is a power of two, these three steps combined together involves $N(1+\log_2 N)$ multiplications and the same number of additions. This should be compared with $N^2$ multiplications and the same number of additions for direct implementation of (\ref{eq:Z}). For a typical value of $N=1024$, this leads to a complexity reduction of 99\%.

\section{USR Traffic and Interference Level}

USR does not require additional spectral resources other than those already allocated for the normal network operation. However, it introduces interference to the other communications in the network. 

The current discussion in 5G NR suggests that each SR can be a single bit that signals ``a particular user requires resources to transmit some information to the BS.'' In USR, by spreading each bit over a ZC sequence, a processing gain in the order of $30$~dB or more is easily attainable. Hence, USRs can be transmitted at a level that minimally affects the thermal/interference noise observed at the BS input. In addition, by adopting an on/off keying strategy where nothing will be transmitted when there is no SR from a UE, the average level of the interference is reduced significantly. To clarify this point, we present a particular use case.

Recent contributions to the development of the 5G NR specification, e.g., \cite{3GPP-R1-1611223}, report use cases where in a cell there might be as many as 20 UEs with each UE, on average, transmitting a maximum of 500 SRs/sec.  With a maximum of $20$ UEs, the total traffic load may be as high as $500\times 20 = 10,000$ SRs per second. With an OFDM symbol rate of about $30,000$ (corresponding to a subcarrier spacing 30 kHz; a numerology that many recent NR contributions have assumed), one will find that only one third of OFDM symbols will be interfered by USRs. This along with the fact that each USR signal introduces a very low level of interference, as reflected in the numerical results that are presented below, leads to a fraction of a decibel increase in the interference level above what already exists in a typical network. We perceive this to be an acceptable investment for the return of low latency in the 5G NR.

\section{Analysis of CP-DSSS Signaling Method}\label{sec:Analysis}

In the case that no information bit is transmitted from user $u$, i.e., $b_{u,i}=0$ for $i=0,1,2,\dots,K$, (13) reduces to
\be
x = c_{i,j}^{(u)}|_{{b_{u,i},b_{u,j}=0}} = | \Re [ \bv_{u,i}^{\rm H} \bv_{u,j} ] | .
\ee
Let $p_0$ denote the probability that the random variable $x$ is greater than a given threshold $\eta$, i.e., 
\be \label{eqn:p0}
p_0 = \Pr( x  > \eta ) .
\ee
For a predetermined value of $p_0$ and assuming that the probability density function (PDF) of the random variable $x$ is known, the threshold $\eta$ can be readily obtained using the above equation. Here, the PDF of $x$ can be found using the
derivations given in the Appendix. The result is
\be \label{eqn:xpdf}
f_X(x) = \frac{\left( \frac{2x}{\sigma_n^2} \right)^{L-\frac{1}{2}} K_{L-\frac{1}{2}}\left( \frac{2x}{\sigma_n^2} \right) }{ \frac{\sigma_n^2}{2} 2^{L-\frac{3}{2}} \pi^\frac{1}{2} \Gamma(L) }, ~~~~~ x \geq 0,
\ee
where $\Gamma(\cdot)$ is the Gamma function and $K_a(\cdot)$ is the $a$-order modified Bessel function of second kind. From (\ref{eqn:xpdf}), we can express the cumulative density function (CDF) of $x$ as $F_X(x) = \int_{0}^x f_X(t)\, dt$. Then, the threshold $\eta$ can be found by finding the root of the non-linear function 
\be \label{eqn:p02}
1-F_X(\eta) - p_0 = 0 .
\ee

In order to utilize the variables $c_{i,j}^{(u)}$ and identify the presence of a packet, we incorporate the following idea. At the BS, if at least $M$ out of $n=\frac{K(K+1)}{2}$ decision variables $c_{i,j}^{(u)}$, for $i\neq j$, are greater than the threshold $\eta$, we say that hypothesis $H_1$ is selected. Here, $M$ is a design parameter.

In order to find the false alarm probability of this algorithm, we note that the random variables $c_{i,j}^{(u)}$ are not necessarily independent. This is due to the fact that for example, $c_{i,j}^{(u)}$ and $c_{i,k}^{(u)}$ have a common constituent variable $\by_{u,i}$. This dependency can complicate the false alarm rate calculations. Here, in order to simplify the derivations, we assume that the random variables $c_{i,j}^{(u)}$, for $i \neq j$, are independent, and evaluate the impact of this assumption on the false alarm rate calculations numerically. According to this assumption, the false alarm rate can be approximated as
\begin{align} \label{eqn:PFA}
P_{\rm FA} &\approx \sum_{i=M}^n {n \choose i} p_0^{i} (1-p_0)^{(n-i)} \nonumber \\
&= I_{p_0} (M,n-M+1) ,
\end{align}
where $I_x(a,b) = \frac{1}{\mathrm{B}(a,b)} \int_{0}^{x}t^{a-1}\,(1-t)^{b-1}\,dt$ is the regularized incomplete beta function, and $\mathrm {B} (a,b)=\int _{0}^{1}t^{a-1}(1-t)^{b-1}\,dt$ denotes the beta function. The second line in (\ref{eqn:PFA}) follows from the fact that the CDF of the binomial distribution $X_{\rm binomial}$ with parameters $n$ and $p$ is given by
\begin{align}
F(k;n,p)&=\Pr(X_{\rm binomial} \leq k) \nonumber \\
&=\sum_{i=0}^k {n \choose i} p^{i} (1-p)^{(n-i)} \nonumber \\
&=I_{1-p}(n-k,k+1) .
\end{align}

We consider a constant false alarm rate (CFAR) detection method, where the probability of false alarm $P_{\rm FA}$ is fixed to a given value. From $P_{\rm FA}$, the value of $p_0$ can be obtained by solving (\ref{eqn:PFA}). This is essentially finding the inverse of the regularized incomplete beta function $I_x(a,b)$ with respect to the integration limit $x$. The MATLAB function \texttt{betaincinv} finds this inverse using the Newton's method. After finding $p_0$, the threshold $\eta$ is found using (\ref{eqn:p02}) as discussed above.

\section{Numerical Results}

\begin{figure}[!t] 
\centering
\subfigure[]{\includegraphics[scale=0.55]{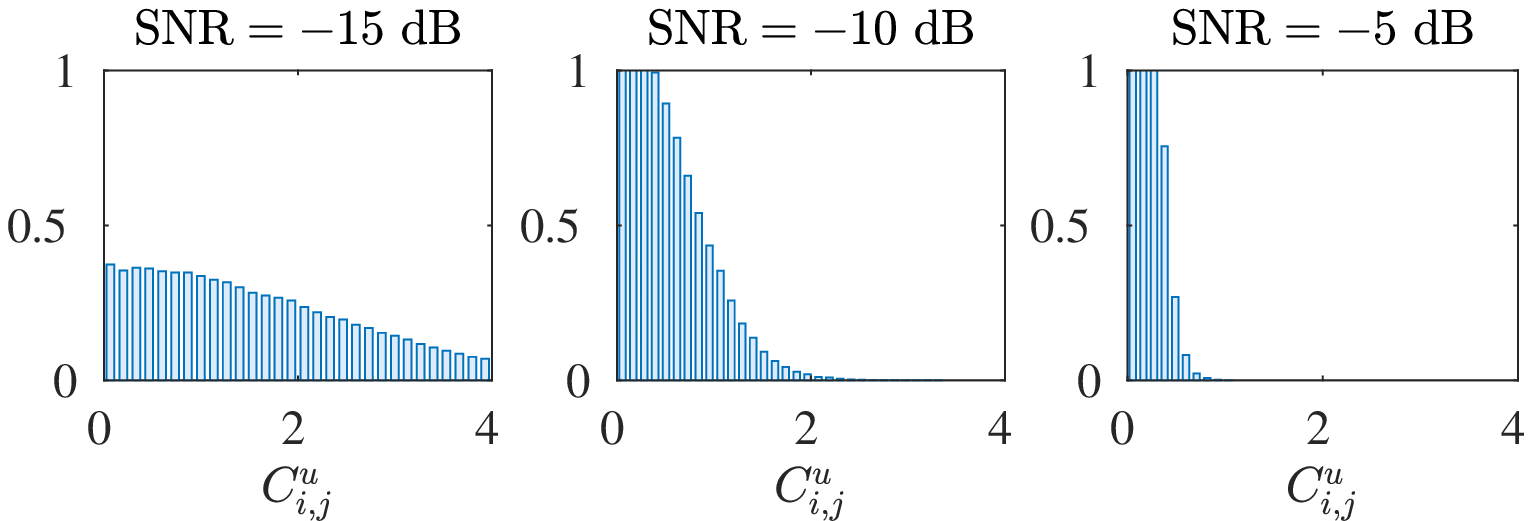}%
\label{fig:dist1}}
\subfigure[]{\includegraphics[scale=0.55]{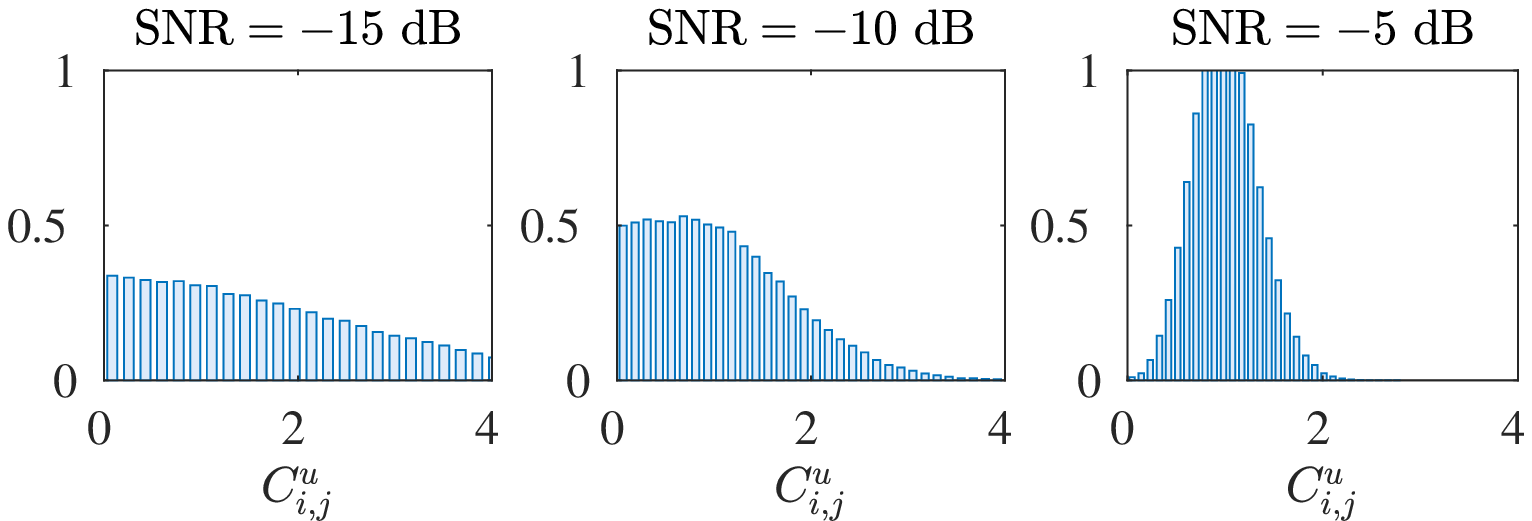}%
\label{fig:dist2}}
\caption{Distribution of the random variable $C_{i,j}^u$, for $i\neq j$, in (a) hypothesis $H_0$ and (b) hypothesis $H_1$. \vspace{-0.1in}}
\label{fig:dist}  
\end{figure}

In this section, we incorporate numerical simulations to show the efficacy of the CP-DSSS signaling method as well as to validate the analysis of the previous sections. To construct the CP-DSSS symbols, we use the ZC sequence length and CP length of $N=1024$ and $72 \approx 0.07 N$, respectively. Moreover, to simulate the multipath fading channel, we utilize the TDL-A channel model with RMS (root mean square) delay spread of $300$ ns. This channel model has been recently proposed in 3GPP, \cite{3gpp.38.900}, for frequency spectrum above 6 GHz. At the receiver side, the channel length of $L=40$ is assumed in the detection process, and the noise variance is estimated simply by measuring the power of the incoming signal noting that the CP-DSSS signal is well below the noise level.

Fig.~\ref{fig:dist} shows the distribution of the random variable $C_{i,j}^u$, for $i \neq j$, under hypotheses $H_0$ and $H_1$. As mentioned in Section \ref{sec:detection}, in the case of hypothesis $H_0$, the distribution of $C_{i,j}^u$ is different from the case of hypothesis $H_1$. In particular, in hypothesis $H_1$, there exist an offset $b_i b_j \| \h_u \|^2$ in the distribution which comes from the first term in (\ref{eq:yui}). This phenomenon is exploited at the receiver for detecting the signal. Moreover, as it can be conveyed from Fig.~\ref{fig:dist}, the distinction between $H_1$ and $H_0$ becomes less evident in very low SNR levels. For example, when SNR $=-15$ dB, the distributions are almost identical in both hypotheses, which suggests the difficulty in signal detection.

\begin{figure}[!t]
\centering
\includegraphics[scale=0.6]{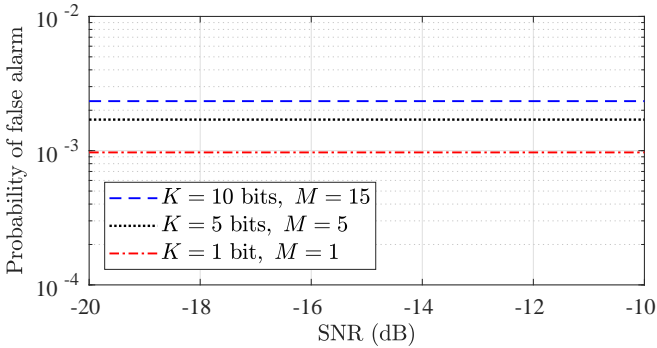} \vspace{-0.1in}
\caption{ Probability of false alarm obtained using simulation. Here, $P_{\rm FA}$ is approximated as $P_{\rm FA} = 0.001$ using the independence assumption.}
\label{fig:PFA} \vspace{-0.15in}
\end{figure}

We now evaluate the approximation made in (\ref{eqn:PFA}) to calculate the probability of false alarm. We fix the theoretical false alarm rate to $P_{\rm FA} = 0.001$ and obtain the threshold $\eta$ according to the derivations in Section \ref{sec:Analysis}. However, we note that in simulations, this threshold leads to a false alarm probability that might be different from the theoretical value, here, $0.001$. Fig.~\ref{fig:PFA} shows the probability of false alarm obtained using simulations for different values of $K$ and $M$. As the figure shows, when $K=1$ bit is transmitted, the theoretical false alarm rate matches perfectly with the simulated one. However, as the number of transmitted bits increases, we observe a small degradation in the false alarm rate due to the approximation in (\ref{eqn:PFA}). For example, when $K=10$, the simulated false alarm rate is $0.0024$.

\begin{figure}[!t]
\centering
\includegraphics[scale=0.6]{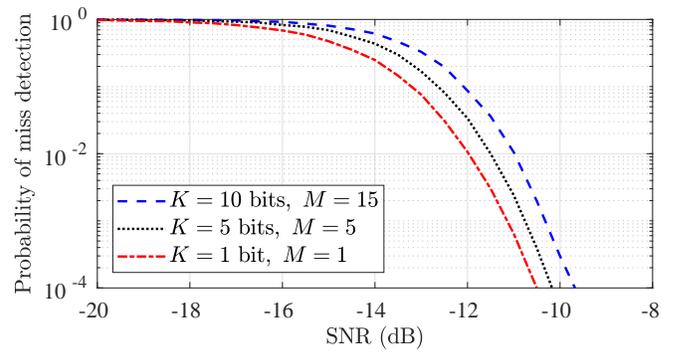} \vspace{-0.1in}
\caption{ Probability of miss detection.  }
\label{fig:PMD}
\end{figure}

Having fixed the false alarm rate of the receiver according to Fig.~\ref{fig:PFA}, we now evaluate the miss detection probability. Fig~\ref{fig:PMD} shows the miss detection rate for different values of $K$ and $M$ and as SNR varies. In this experiment, the threshold $\eta$ is obtained by fixing the theoretical false alarm rate to $P_{\rm FA} = 0.001$ for all SNR values. According to this figure, as the number of transmitted bits $K$ increases, the curve corresponding to the miss detection probability drifts to right by a small amount. For example, when $K$ is increased from 1 to 10, the curve is shifted to the right by less than 1 dB when $P_{\rm MD} = 0.001$.

\begin{figure}[!t]
\centering
\includegraphics[scale=0.58]{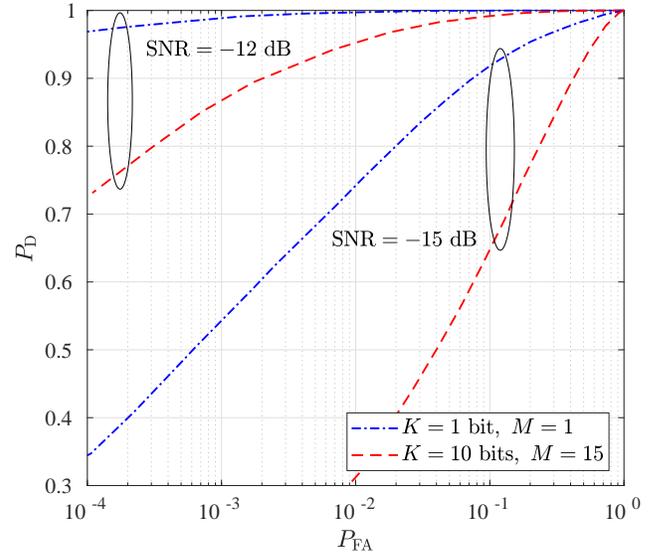} \vspace{-0.1in}
\caption{ Receiver operating characteristic (ROC). } 
\label{fig:ROC} \vspace{-0.15in}
\end{figure}

As the above experiments illustrate, when the number of transmitted bits $K$ increases, both $P_{\rm FA}$ and $P_{\rm MD}$ undergo a small degradation. In order to show the effect of increasing $K$ on both $P_{\rm FA}$ and $P_{\rm MD}$ together, we show the receiver operating characteristic (ROC) curve. Fig.~\ref{fig:ROC} shows the ROC curve for different values of $K$ and $M$ and two choices of SNR. According to this figure, when $K=1$ bit transmission is considered, false alarm probability of $0.001$ and detection probability of $0.99$ can be obtained at the SNR level of $-12$ dB. On the other hand, for the $K=10$ bit transmission scenario, a higher SNR level is required to achieve that performance.

\begin{figure}[!t]
\centering
\includegraphics[scale=0.6]{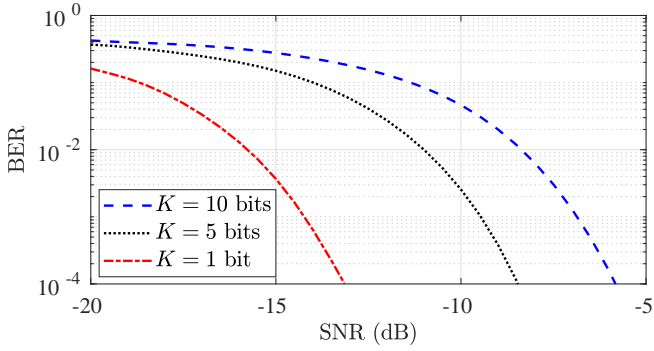}
\caption{Uncoded BER for different values of $K$.} \vspace{-0.15in}
\label{fig:BER}
\end{figure}

We next evaluate the efficacy of information recovery scheme given in Section \ref{sec:recovery}. Fig.~\ref{fig:BER} shows the \emph{uncoded} bit error rate (BER) assuming that the CP-DSSS signal is successfully detected. As the number of information bits increases, assuming that the total transmitted power is fixed, the power portion corresponding to the ZC sequence of each bit becomes smaller. This affects the information recovery accordingly. For example, when the number of bits is increased from $K=1$ to $K=10$, the uncoded BER curve is shifted $10\log_{10} \left( \frac{10+1}{1+1} \right) = 7.4$ dB to the right. Here, we have accounted for the reference bit as well as the additional information bits. 

\section{Conclusion}

An underlay control signaling method for URLLC applications was presented and its performance was analyzed. The proposed design is a special form of direct sequence spread spectrum which makes use of a CP to allow simple synchronization with OFDM symbols within an LTE network. Zadoff-Chu sequences were used for spreading and the signal design was made such that one or multiple bits could be transmitted over each control packet which has the same length as OFDM symbols. The design is also such that the users can be identified by the position of the control bits that they transmit within each OFDM symbol interval. Accordingly, the presence or absence of a user signal is identified by the presence or absence of the control bits at the respective positions. This feature of our design is very import, as those users that have no request for scheduling send nothing and, thus, do not cause any interference to the normal communications in the network.

\appendix

In order to derive the distribution of the random variable $x$, we start from the random variable $\tilde{x} = \Re [ \bv_{u,i}^{\rm H} \bv_{u,j} ]$. We note that $\tilde{x}$ is a summation of $2L$ real-valued random variables, each of them being the multiplication of two Gaussian random variables. All the random variables are independent of one another and have the mean of zero and variance of $\frac{\sigma_n^2}{2}$. To find the PDF of $\tilde{x}$, we first recall that the multiplication of a pair of independent zero mean Gaussian random variable $p$ and $q$ is a new random variable with the following distribution:
\be
pq \sim \frac{1}{\pi \sigma_p \sigma_q} K_0\left(\frac{|pq|}{\sigma_p \sigma_q}\right) ,
\ee
where $K_0(\cdot)$ is zero-order modified Bessel function of second kind. We thus note that the random variable $\tilde{x}$ may be written as $\tilde{x} = \sum_{i=0}^{2L-1} x_i$,
with  
\be \label{eqn:xi}
x_i \sim \frac{1}{\frac{\pi \sigma_n^2}{2}} K_0 \left(\frac{|x_i|}{\frac{\sigma^2_n}{2}} \right) = \frac{2}{\pi \sigma^2_n} K_0 \left( \frac{2|x_i|}{\sigma^2_n} \right) .
\ee
Direct calculation of the PDF of $\tilde{x}$ from above requires convolution of $2L$ functions of the form (\ref{eqn:xi}). This is very complex to evaluate. An easier method is to first note that the random variable $x_i$ has the characteristic function
\be
\phi_{x_i}(t) = \frac{\frac{2}{\sigma_n^2}}{\left( t^2 + \frac{4}{\sigma_n^4} \right)^{\frac{1}{2}}} .
\ee
Also, (\ref{eqn:xi}) implies that 
\be
\phi_x(t) = \prod_{i=0}^{2L-1} \phi_{x_i}(t) = \frac{\left(\frac{2}{\sigma^2_n}\right)^{2L}}{\left( t^2+\frac{4}{\sigma^4_n} \right)^L} .
\ee
Converting this characteristic equation to its associated PDF, leads to 
\be 
f_{\tilde{X}}(\tilde{x}) = \frac{\left( \frac{2 \tilde{x}}{\sigma_n^2} \right)^{L-\frac{1}{2}} K_{L-\frac{1}{2}}\left( \frac{2\tilde{x}}{\sigma_n^2} \right) }{ \frac{\sigma_n^2}{2} 2^{L-\frac{3}{2}} \pi^\frac{1}{2} \Gamma(L) } .
\ee
This leads to the distribution of the desired random variable $\tilde{x} = |x|$ as given in (\ref{eqn:xpdf}).

\bibliographystyle{IEEEtran}
\bibliography{bibfile.bib}

\end{document}